\DeclareUnicodeCharacter{202F}{\,}
\documentclass{article}
\usepackage{spconf,amsmath,graphicx}

\usepackage[colorlinks=true,linkcolor=blue,citecolor=blue,urlcolor=blue]{hyperref}

\usepackage{url}
\usepackage{graphicx}
\usepackage{soul}
\usepackage{booktabs}
\usepackage{bm}
\usepackage{multirow}

\usepackage{pifont}

\usepackage{xcolor}
\usepackage{booktabs}

\usepackage{subcaption}
\usepackage{graphicx}
\usepackage[normalem]{ulem}
\usepackage{tabularx}
\usepackage{amssymb}
\usepackage{colortbl}
\usepackage{enumitem}
\usepackage{amsmath}

\usepackage{orcidlink}
\newcolumntype{L}{>{\raggedright\arraybackslash}X}  

\definecolor{colMAEST}{HTML}{30123B}
\definecolor{colCLAP}{HTML}{4685FA}
\definecolor{colWhisper}{HTML}{1AE4B6}
\definecolor{colOMARRQ}{HTML}{A4FC3C}
\definecolor{colMERT}{HTML}{E4450A}
\definecolor{colMusicFM}{HTML}{7A0403}

\newcommand{\modelos}[1]{{\fontsize{8.6pt}{9.4pt}\selectfont\textsc{#1}}}

\name{
Pedro Ramoneda$^{1}$\thanks{Corresponding author: \url{pedro.ramoneda@upf.edu}},\ 
Pablo Alonso-Jiménez$^{1}$,\ 
Sergio Oramas,\ 
Xavier Serra$^{1}$,\ 
Dmitry Bogdanov$^{1}$
}
\address{
$^{1}$Music Technology Group, Universitat Pompeu Fabra, Barcelona, Spain
}

\title{Benchmarking Music Autotagging with MGPHot\\ Expert Annotations vs. Generic Tag Datasets}

\begin{document}

\ninept
\maketitle


\begin{abstract}
Music autotagging aims to automatically assign descriptive tags, such as genre, mood, or instrumentation, to audio recordings. 
Due to its challenges, diversity of semantic descriptions, and practical value in various applications, it has become a common downstream task for evaluating the performance of general-purpose music representations learned from audio data.
We introduce a new benchmarking dataset based on the recently published MGPHot dataset, which includes expert musicological annotations, allowing for additional insights and comparisons with results obtained on common generic tag datasets. While MGPHot annotations have been shown to be useful for computational musicology, the original dataset neither includes audio nor provides evaluation setups for its use as a standardized autotagging benchmark.
To address this, we provide a curated set of YouTube URLs with retrievable audio, and propose a train/val/test split for standardized evaluation, and precomputed representations for seven state-of-the-art models.
Using these resources, we evaluated these models in MGPHot and standard reference tag datasets, highlighting key differences between expert and generic tag annotations. Altogether, our contributions provide a more advanced benchmarking framework for future research in music understanding.
\end{abstract}


\section{Introduction}

Music autotagging aims to derive rich semantic descriptors, such as genre, mood, instrumentation, rhythm, harmony, production, and composition traits, directly from raw audio~\cite{bertin2011automatic,duan2014survey,marques2011three}. Such an analysis has great potential in various applications, especially in music streaming and recommendation services and in the management of music catalogs, where automatic audio understanding helps organize, filter, and personalize content.
The standard way to evaluate music autotagging models uses a two-step pipeline~\cite{mccallum2022supervised}. In this setup, a shallow model is trained on top of the output of a pretrained representation model (audio encoder). This approach is simple and efficient because it is lightweight and allows reuse of representations across multiple tasks. First, models are trained on large audio datasets to learn music representations. Second, a small discriminative head is trained for each downstream tagging task. 
This type of pipeline is highly versatile, supporting a wide range of downstream tasks beyond autotagging, and has achieved strong results in multiple Music Information Retrieval (MIR) applications
~\cite{ma2024foundation}. 
However, current evaluation practices for general-purpose music representations are limited, and there is a need for rigorous, well-designed evaluation benchmarks, as performance can vary greatly depending on the research datasets and metrics used.

Although early studies used small datasets like \emph{GTZAN}\cite{tzanetakis2002musical} and \emph{Latin Music Database}~\cite{silla2008latin}, their size and taxonomies proved insufficient for robust evaluation~\cite{bogdanov2016crosscollection,sturm2013gtzan,sturm2015faults}.
Currently, researchers rely on larger crowdsourced datasets with generic tag annotations, such as \emph{MagnaTagATune}~\cite{law2009evaluation} which covers around 5\,000 songs from an independent record label, and \emph{MTG‑Jamendo}~\cite{bogdanov2019mtg}, which compiles more than 50\,000 amateur-produced songs. However, these annotations have been found to be inconsistent and have varying reliability, which hinders fine-grained model evaluation.

The recently introduced \emph{MGPHot} dataset provides expert musicological annotations for 21{,}320 tracks from the \emph{Billboard Hot 100} between 1958 and 2022. Each track is annotated with 58 continuous attributes grouped into seven categories: rhythm, compositional focus, harmony, instrumentation, sonority, vocals, and lyrics, curated by professional musicians from the Music Genome Project \cite{oramas2025mgphot}. 
Notably, these characteristics are different and more detailed than the tags previously used in research, e.g., ``Vocal Grittiness'', ``Harmonic sophistication'', or ``Aural Intensity'', instead of common labels such as ``Vocal'' or genre tags, which can offer new perspectives for evaluating music understanding. The creators of the dataset demonstrated
how these annotations can be used to analyze musical trends~\cite{oramas2025mgphot}. However, the distribution of the dataset comprises tracks metadata and visualizations; no audio files or canonical evaluation splits are provided, which prevents the use of the dataset in research involving audio-based models.


In this work, we propose using the \emph{MGPHot} dataset to benchmark music audio representation models by matching its tracks to audio from YouTube. We retrieve all tracks, 56.43\% from official sources, such as artist-topic channels and label uploads, and define the first canonical train/val/test split for \emph{MGPHot}, together with the derived tag annotations.

We evaluated seven state-of-the-art models, \modelos{Whisper}~\cite{radford2023robust}, \modelos{CLAP}~\cite{laionclap2023}, \modelos{MAEST}~\cite{alonso2023efficient}, \modelos{MERT}~\cite{li2024mert}, \modelos{MusicFM}~\cite{wonz2024}, and \modelos{OMAR-RQ}~\cite{alonso2025omar}, on the \emph{MGPHot}~\cite{oramas2025mgphot}, \emph{MTG‑Jamendo}\cite{bogdanov2019mtg}, and \emph{MagnaTagATune}~\cite{law2009evaluation} datasets. 
For more detailed insights, we also map the generic tag vocabularies of \emph{MTG‑Jamendo} and \emph{MagnaTagATune} into higher‑level musical categories.

\noindent Contributions:

\begin{itemize}[leftmargin=1.2em, labelsep=0.4em, itemsep=0pt, topsep=0pt, parsep=0pt, partopsep=0pt]
\item Extended metadata for the expert-annotated \emph{MGPHot} dataset, including curated YouTube URLs, code, canonical train/val/test splits, tag set, and pre-extracted features from seven state-of-the-art models.
\item Benchmark comparison of seven leading self-supervised representation models in \emph{MGPHot}, \emph{MTG-Jamendo}, and \emph{MagnaTagATune}, including a per-category evaluation across musical feature groups.
\item Although all evaluated models claim state-of-the-art performance, our benchmark reveals ranking shifts across datasets, categories, and tags. This provides a clear picture of the current state of music autotagging and highlights the critical role of extensive cross-dataset and category evaluation.
\end{itemize}

Altogether, these resources and findings promote more rigorous evaluation practices 
for future research on music representation learning systems.

\section{Gathering audio for \emph{MGPHot}}
\label{sec:audio_compilation}

\begin{figure}[t]
  \centering
  \includegraphics[width=0.95\columnwidth]{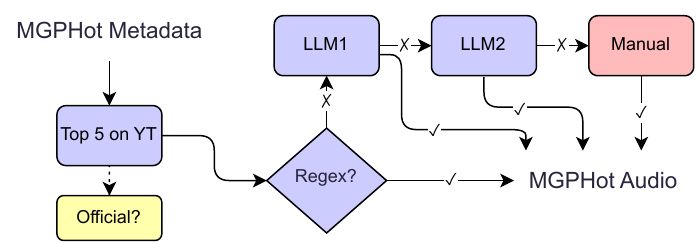}
  \caption{Pipeline for compiling the \emph{MGPHot} audio archive.
    Percentages indicate the contribution of each step.}
  \label{fig:mgphot_audio_flow}
\end{figure}

Figure~\ref{fig:mgphot_audio_flow} illustrates the pipeline we followed to collect YouTube URLs for the metadata of \emph{MGPHot}.
We started from the metadata for the 21,320 chart tracks. For each track, we searched YouTube using the title of the song and the artist's name, keeping the top five results.
A regular expression match between the track title and the video title yielded a direct hit in 72.91\% of the cases.
When the match failed, we applied two large language model (LLM) iterations with QWEN2.5 \_32B~\cite{qwen2.5}:
the first compared only titles and artist information, adding 22.86\% matches, while the second also examined video descriptions and resolved another 739 tracks (3.47\%), leaving only 163 tracks for manual verification.
In parallel, we checked whether each video came from an official artist channel, confirming that 56.43\% of the final matches are official uploads.
This procedure linked all \emph{MGPHot} tracks with YouTube while minimizing the ratio of unofficial sources.

We distribute YouTube URLs and metadata, along with a script for local audio downloads in a reproducible manner.
Because the original dataset license forbids redistribution of derivative files, we
avoid distributing the original annotations.
Instead, we provide a script that downloads the official annotations from Zenodo, merges them with our YouTube metadata, and rebuilds the canonical subsets.
MD5 checksums are included to ensure the integrity and canonical formatting of the reconstructed files.

We organize the annotations into two supervised tasks or subsets:
\emph{MGPHot-reg} retains the $58$ original continuous values in the range $[0, 1]$;
\emph{MGPHot-tag} discretizes these values into three categorical tags, corresponding to the intervals ``Low'' $(0,\,0.33)$, ``Moderate'' $[0.33,\,0.66)$, and ``High'' $[0.66,\,1]$. 
These categories account for $12.0\%$, $55.5\%$, and $32.5\%$ of the total tags, respectively. Note that, except for ``Major/Minor'', the value $0$ is skipped because it corresponds to no tag.
Some descriptors do not exhibit values within all intervals, resulting in a total of 174 distinct tags.
Both subsets use the train/val/test partitions from Section~\ref{sec:dataset_partitioning}. 

All extended metadata are released under the CC BY--NC 4.0 license.\footnote{\href{https://creativecommons.org/licenses/by-nc/4.0/}{https://creativecommons.org/licenses/by-nc/4.0/}}
The subsets, audio download and reconstruction scripts are available in a public GitHub repository.\footnote{\href{https://github.com/MTG/MGPHot-audio}{https://github.com/MTG/MGPHot-audio}}
The audio embeddings evaluated in this paper, along with the per-category tags for \emph{MTG-Jamendo} and \emph{MagnaTagATune} datasets, are available on Zenodo.\footnote{\href{https://doi.org/10.5281/zenodo.16993068}{https://doi.org/10.5281/zenodo.16993068}}
These embeddings facilitate autotagging evaluation by allowing researchers to train lightweight classifiers on the same features without redownloading audio or rerunning feature extraction. They can be used both to replicate our probing protocol exactly and to evaluate new music understanding models under alternative protocols or classifiers.

\begin{figure}[t]
  \centering
  \includegraphics[width=\columnwidth]{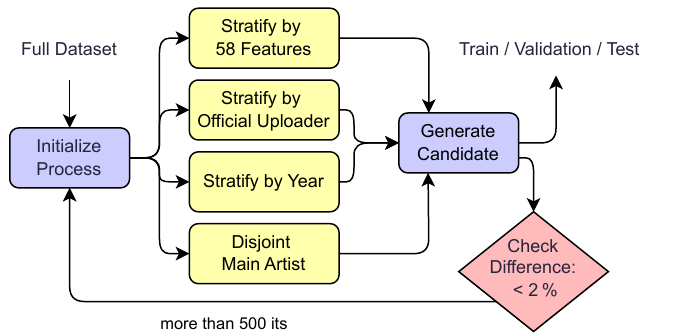}
  \caption{Flowchart of the split‑generation procedure for \emph{MGPHot}.}
  \label{fig:mgphot_split_flow}
\end{figure}

\section{MGPHot Dataset Partitioning}
\label{sec:dataset_partitioning}

Figure~\ref{fig:mgphot_split_flow} sketches the automatic procedure 
used to create the canonical train/val/test split for \emph{MGPHot}.  
We start from the full collection. For conducting the iterative split generation, each candidate split must satisfy four constraints:

\begin{itemize}[leftmargin=1.2em, labelsep=0.4em, itemsep=0pt, topsep=0pt, parsep=0pt, partopsep=0pt]
\item \textit{Stratification by the 58 expert descriptors.}  
      We match the marginal distribution of every descriptor across the three sets. 
\item \textit{Balanced official uploads.}  
      The ratio of videos from official artist channels is kept similar in all sets.
\item \textit{Balanced year.}  
      The original study~\cite{oramas2025mgphot} stresses the significance of the song's release year. The proportion of years is consistently maintained across all groups.
\item \textit{Disjoint main artists.}  
      Tracks of the same main artist appear in only one set.
\end{itemize}

We tested random splits using artist–disjoint multilabel stratification until the maximum absolute difference in label proportions between each set and the overall distribution fell below $2\%$ (computed over all label bins). The resulting split is released together with the extended metadata.

\begin{table}[t]
\small
\centering
\begin{tabularx}{\columnwidth}{l c c c c c}
\toprule
  Dataset & Type & Tags . & Samples & Avg.~Tags \\
\midrule
\textit{MagnaTagATune}   & bin. & 188 & 5\,405 & 3.46 \\[4pt]

\textit{MTG-Jamendo}   & bin. & 195  & 55\,701 & 4.18 \\[4pt]

\textit{MGPHot-reg}  & cont.& 58 & 21\,320 & 58 \\[4pt]
\textit{MGPHot-tag}  & bin. & 174 &  21\,320 & 58 \\
\bottomrule
\end{tabularx}
\caption{
Overview of the datasets used in this study. 
\textit{bin.}: binary tags; \textit{cont.}: continuous annotations; 
\textit{Avg.~Tags}: descriptor density (average active tags per track). 
\emph{MGPHot} is provided in two variants: regression (\textit{reg}) with continuous descriptors and autotagging (\textit{tag}) with binarized labels.
}
\label{tab:data_overview}
\vspace{-0.3cm}
\end{table}

\section{Evaluation Protocol}
\label{sec:evaluation_protocol}

\textbf{Dataset splits.}
We follow the train/validation/test partition used in previous work for ~\emph{MagnaTagATune} ~\cite{wonz2024, alonso2025omar}, and the \emph{split 0} base autotagging partition for \emph{MTG--Jamendo}.
For \emph{MGPHot} we use our proposed split.

\textbf{Tasks.}
We consider two tagging settings.
For \emph{MagnaTagATune}, \emph{MTG--Jamendo}, and \emph{MGPHot-tag}, we perform multilabel classification with sigmoid output and binary cross entropy.
For \emph{MGPHot-reg}, we perform regression of 58 continuous descriptors in $[0,1]$ with mean squared error and without sigmoid.

\textbf{Probe architecture.} For each pretrained encoder, we freeze the encoder and attach a two--layer MLP (512 hidden units) with ReLU. The probe uses one vector per track, obtained by mean pooling over time, a standard choice in music autotagging. We train with AdamW (lr \(3\times10^{-4}\), wd \(10^{-2}\)), batch size \(128\), and early stopping (patience \(50\)).

\textbf{Audio encoders.}
We evaluated seven pretrained audio encoders, as shown in Table~\ref{tab:models_overview}, selected for their relevance and reported strong performance.
\modelos{MAEST} is a bidirectional transformer trained with a music style classification objective on a large audio collection annotated by Discogs genre metadata~\cite{alonso2023efficient}.  
\modelos{CLAP} has text and audio encoders trained with contrastive loss to align paired audio-text examples. The text is natural language metadata: captions, titles, and tags that describe sources, instrument, genre, mood, or sound events~\cite{laionclap2023}.
\modelos{Whisper} features an encoder-decoder transformer architecture and is trained for automatic speech recognition in several languages~\cite{radford2023robust}.
Finally, we consider three self-supervised audio masked language modeling models following different tokenization approaches.
While \modelos{MERT} targets a combination of tokens derived from RVQ and CTQ clusters~\cite{li2024mert}, \modelos{MusicFM} creates target tokens by applying random codebook quantization over mel spectrograms~\cite{wonz2024}, and \modelos{OMAR-RQ} adopts a version that extends this approach to a multilabel setting using multiple codebooks in parallel~\cite{alonso2025omar}.

\begin{table}[t]
\small
\centering
\setlength\tabcolsep{4pt}
\begin{tabularx}{\columnwidth}{l c c c l}
\toprule
\textbf{Model} & \textbf{Task} & \textbf{Hours} & \textbf{$\theta$ (M)} & \textbf{Architecture} \\
\midrule
\modelos{Whisper}    & ASR                 & 680{,}000 & 635     & Transformer \\
\modelos{CLAP}       & text/audio CL       & 4{,}325   & 31    & HTS-AT \\
\modelos{MAEST}      & genre prediction      & 330{,}000 & 86     & Transformer \\
\modelos{MERT}       & MATP                & 160{,}000 & 330    & Transformer \\
\modelos{MusicFM}    & MATP                & 8{,}000   & 330    & Conformer \\
\modelos{OMAR-RQ}    & MATP                & 330{,}000 & 580    & Conformer \\
\bottomrule
\end{tabularx}
\caption{Overview of seven music/audio encoders. ``$\theta$'' = millions of trainable parameters.
Tasks: Automatic Speech Recognition (ASR), Contrastive Learning (CL), masked audio token prediction (MATP). 
}
\label{tab:models_overview}
\vspace{-0.3cm}
\end{table}

\begin{table*}[ht!]
\centering
\begin{tabular}{lcccc}
\toprule
\textbf{Model} & \textbf{MagnaTagATune} & 
\textbf{MTG-Jamendo} & 
\textbf{MGPHot-tag} &
\textbf{MGPHot-reg} \\
& MAP~$\uparrow$ & MAP~$\uparrow$ 
& MAP~$\uparrow$
& RMSE~$\downarrow$ \\
\midrule
\modelos{Whisper}~\cite{radford2023robust}   
& 0.376 $\pm$ 0.000
& 0.099 $\pm$ 0.001

& \cellcolor{gray!20}0.365 $\pm$ 0.001
& \cellcolor{gray!20}0.167 $\pm$ 0.000\\

\modelos{CLAP}~\cite{laionclap2023}     
& 0.443 $\pm$ 0.000
& 0.124 $\pm$ 0.000

& \cellcolor{gray!20} \textbf{\uline{0.375 $\pm$ 0.000}}
& \cellcolor{gray!20} 0.165 $\pm$ 0.000\\

\modelos{MAEST}~\cite{alonso2023efficient}
& \cellcolor{gray!20}\textbf{\uline{0.493 $\pm$ 0.001}}
& \cellcolor{gray!20}\textbf{\uline{0.154 $\pm$ 0.004}}
& 0.347 $\pm$ 0.000
& 0.172 $\pm$ 0.000\\


\modelos{MERT}~\cite{li2024mert}       
& 0.442 $\pm$ 0.002
& \cellcolor{gray!20}0.139 $\pm$ 0.001
& \cellcolor{gray!20} 0.365 $\pm$ 0.002
& \cellcolor{gray!20} \textbf{0.164 $\pm$ 0.001}\\

\modelos{MusicFM}~\cite{wonz2024}    
& \cellcolor{gray!20}0.444 $\pm$ 0.000
& 0.122 $\pm$ 0.000

& 0.358 $\pm$ 0.000
& 0.172 $\pm$ 0.001\\

\modelos{OMAR-RQ}~\cite{alonso2025omar}     
& \cellcolor{gray!20}0.484 $\pm$ 0.001
& \cellcolor{gray!20}0.135 $\pm$ 0.001

& \cellcolor{gray!20} 0.365 $\pm$ 0.001
& 0.171 $\pm$ 0.001\\
\bottomrule
\end{tabular}
\caption{Model performance across four tasks. Metrics (macro over tags): MAP for classification and RMSE for regression. The best result is marked in bold and underlined if the improvement is significant over the second best according to a paired two-tailed Student’s t-test ($p<0.05$).
The top-3 per column appear with a light gray background.} 
\label{tab:results_table}
\end{table*}

\section{Results}

Table~\ref{tab:results_table} reports the mean average precision (MAP~$\uparrow$) for the three tagging tasks and root mean-squared error (RMSE~$\downarrow$) for the regression task.\footnote{Chosen for interpretability, MAE and MSE results are available online.}  Each score is the mean of five runs initialized with different seeds.

No single encoder leads in all settings, and differences between models are often limited even when statistically significant. Considering encoders with audio self-supervision, \modelos{MERT} and \modelos{OMAR-RQ} rank consistently among the top models across the four benchmarks, reflecting the strong potential of masked audio token prediction approaches.
Among the approaches with metadata supervision, \modelos{MAEST} leads the two generic tag datasets (\emph{MagnaTagATune} and \emph{MTG-Jamendo}) but performs below par in the \emph{MGPHot} dataset, which has more specific musical features.
\modelos{CLAP} achieves the best results in \emph{MGPHot-tag} and ranks second in \emph{MGPHot-reg}, with no statistically significant difference from \modelos{MERT}, the top model.

\begin{figure*}[t]
  \centering
  \includegraphics[width=\textwidth,clip]{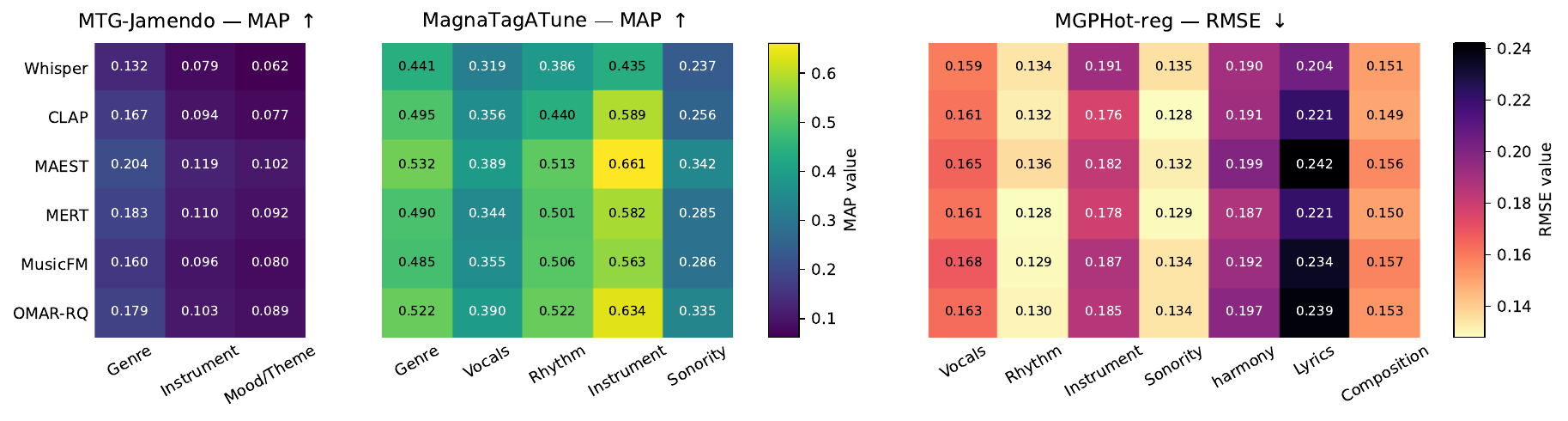}
  \vspace{-0.8cm}
  \caption{Heatmaps with models in rows and categories in columns. 
The two left panels show MAP~$\uparrow$ for \emph{MTG--Jamendo} and \emph{MagnaTagATune} (shared color scale). The right panel shows RMSE~$\downarrow$ for \emph{MGPHot} (separate scale). }
  \label{fig:bars}
\end{figure*}

Figure~\ref{fig:bars} shows how the seven encoders score in each tag category. In both generic tag datasets, \modelos{MAEST} clearly leads, especially in ``Genre'' for \emph{MTG-Jamendo},
\footnote{Note that we use all the tags available for each category, which does not match the official genre, mood/theme, and instrument MTG-Jamendo splits.}
likely due to alignment with its supervised pretraining objective. \modelos{OMAR-RQ} usually follows a few points behind. The heatmap on the right reports the RMSE on \emph{MGPHot}, where smaller values indicate better performance. The differences are relatively small: \modelos{CLAP} achieves the lowest error for ``Instrument'', ``Sonority'', and ``Composition'', and \modelos{MERT} slightly outperforms others in ``Harmony''. 
Interestingly, \modelos{Whisper}, trained in speech, performs poorly in autotagging on generic datasets but is in the top 3 on \emph{MGPHot-reg} and \emph{MGPHot-tag}, due to its high performance for ``Vocals'' and ``Lyrics'' categories, revealed in the analysis of results per category. The difficulty of the category varies between datasets. In the first set, ``Genre'' is the easiest. In \emph{MagnaTagATune}, ``Instrument'' is easier. In \emph{MGPHot}, disparities are larger: ``Lyrics'' is the most challenging, followed by ``Harmony'' and ``Instrument''. Note that even when categories share the same name, results differ substantially because the underlying tags are different across datasets.


We report the performance per tag in an interactive online tool.\footnote{Results per Tag: \href{https://pramoneda.github.io/tagbenchmark}{https://pramoneda.github.io/tagbenchmark}} Performance also varies widely between datasets and tags. For example, tags such as ``piano'' yield similar results across models, whereas others like ``synth'' show stronger differences. We also observe tags that are particularly challenging; in \emph{MGPHot}, lyric-related tags, the ``Major/Minor'' value, and ``Focus on Riffs'' are especially hard.

\section{Discussion and Limitations}
\label{sec:discussion}

Although all encoders considered claim state-of-the-art performance, our study finds no model that consistently leads across all settings. \modelos{MAEST} achieves the best scores in the two generic tag datasets, \modelos{CLAP}, \modelos{Whisper}, and \modelos{MERT} share the top position in detailed musical features annotated by experts and \modelos{OMAR-RQ} remains competitive in all cases. This distribution of winners indicates that there is no single reliable choice.

The results exhibit substantial performance variability across datasets, reflecting the heterogeneity of real-world audio sources and annotations. Findings that hold in \emph{MGPHot}, with refined expert-annotated labels, do not necessarily generalize to generic tag annotations and vice versa. This variability underscores the limitations of evaluations in previous studies, which cannot draw definitive conclusions from generic tag datasets.

Supervised pretraining excels when the downstream tags are aligned with the pretraining labels, as \modelos{MAEST} demonstrates in MTG-Jamendo and MagnaTagATune, which have a large number of genre tags (87 of 185 and 14 of 50, respectively) with the \modelos{MAEST} pretraining set. In contrast, \emph{MGPHot} focuses on other aspects less associated with the musical genre, resulting in a substantial drop in performance. \modelos{CLAP}, which aligns audio with text and operates in a broader semantic space, handles this mismatch better. Meanwhile, masked token audio prediction models trained solely on audio without any metadata supervision provide a balanced trade-off: they do not achieve the best performance, but remain decently competitive. 

The results for tagging and regression on \emph{MGPHot} are broadly similar, but each approach has its advantages. \emph{MGPHot-tag} aligns with how autotagging is commonly
addressed as a classification problem, allowing a direct comparison with previous work. In contrast, regression 
benefits from
the original continuous annotations without 
adding
discretization noise.

Moreover, the lack of consistent improvements from larger models or more data (Table~\ref{tab:models_overview} vs Table~\ref{tab:results_table}) highlights the importance of efficient and sustainable audio encoder design~\cite{holzapfel2024green}.

A limitation of this study is that we only evaluate frozen encoders.
Although full fine-tuning or parameter-efficient updates could raise performance, freezing provides a controlled setting to assess the intrinsic representation quality. 
In addition, our evaluation is restricted to track-level autotagging, continuous or discrete. However, the same encoders could be reused for 
other MIR tasks, such as onset detection, beat tracking, or source separation,
covering a broader scope of music understanding.

\section{Conclusion}
In this paper, we evaluate state-of-the-art music audio representations in music autotagging tasks, using two common generic tag datasets and a new \emph{MGPHot} dataset, which we extend and propose as a new benchmark for audio-based evaluations. The results reveal performance inconsistencies across datasets, highlighting the limitations of relying solely on generic tag datasets in previous studies and underscoring the need for datasets with more detailed annotations and richer insights into different aspects of music description. We release the extended metadata for the \emph{MGPHot} dataset to facilitate further research.

\section*{Acknowledgments}

This work is supported by “IA y Música: Cátedra en Inteligencia Artificial y Música” (TSI-100929-2023-1) funded by the Secretaría de Estado de Digitalización e Inteligencia Artificial and the European Union-Next Generation EU, under the program Cátedras ENIA. We thankfully acknowledge the computer resources at MareNostrum and the technical support provided by Barcelona Supercomputing Center (IM-2024-2-0034).

\bibliographystyle{IEEEtran}
\bibliography{main}

\end{document}